\documentclass[aoas,MSNbibl,nameyear,dvips]{arximspdf}
\usepackage{graphicx}
\doi{10.1214/11-AOAS482}
\volume{5}
\issue{4}
\pubyear{2011}
\firstpage{2425}
\lastpage{2447}

\makeatletter

\newcommand{\btheta}{\bolds\theta}
\newcommand{\bbeta}{\bolds\beta}
\newcommand{\balpha}{\bolds\alpha}
\newcommand{\bx}{\mathbf{x}}
\newcommand{\bs}{\mathbf{s}}
\newcommand{\bu}{\mathbf{u}}

\makeatother

\begin{document}
\begin{frontmatter}

\title{A class of covariate-dependent spatiotemporal covariance
functions for the analysis of daily ozone concentration\thanksref{T1}}
\runtitle{Covariate-dependent spatiotemporal covariance functions}

\thankstext{T1}{Supported by
the Statistical and Applied Mathematical Sciences Institute NSF
Grant DMS-06-35449.}

\begin{aug}
\author[A]{\fnms{Brian J.} \snm{Reich}\corref{}\ead[label=e1]{reich@stat.ncsu.edu}},
\author[B]{\fnms{Jo} \snm{Eidsvik}},
\author[C]{\fnms{Michele} \snm{Guindani}},
\author[D]{\fnms{Amy J.} \snm{Nail}\ead[label=e2]{amynailstat@gmail.com}}
and
\author[E]{\fnms{Alexandra M.} \snm{Schmidt}}
\runauthor{B. J. Reich et al.}
\affiliation{North Carolina State University,
Norwegian University of Science and Technology,
MD Anderson Cancer Center,
Duke University and Universidade Federal do Rio de Janeiro, Brazil}
\address[A]{B. J. Reich\\
North Carolina State University\\
2501 Founders Drive, Box 8203\\
Raleigh, North Carolina 27695\\
USA\\
\printead{e1}} 
\address[B]{J. Eidsvik\\
Department of Mathematical Sciences\\
Norwegian University of Science\\
\quad and Technology\\
7491 Trondheim\\
Norway}
\address[C]{M. Guindani\\
University of Texas\\
\quad MD Anderson Cancer Center\\
1515 Holcombe Blvd.\\
Unit 1411\\
Houston, Texas 77030\\
USA}
\address[D]{A. Nail\\
President, Honestat\\
919-838-9532\\
1105 Somerset Rd\\
Raleigh, North Carolina 27610\hspace*{29pt}\\
USA\\
\printead{e2}}
\address[E]{A. M. Schmidt\\
Universidade Federal do Rio de Janeiro\\
Instituto de Matem\'{a}tica\\
Departamento de M\'{e}todos Estat\'{\i}sticos\\
Caixa Postal 68530\\
CEP.: 21945-970\\
Rio de Janeiro\\
Brazil}
\end{aug}

\received{\smonth{5} \syear{2010}}
\revised{\smonth{4} \syear{2011}}

%
\begin{abstract}
In geostatistics, it is common to model spatially distributed phenomena
through an underlying stationary and isotropic spatial process.
However, these assumptions are often untenable in practice because of
the influence of local effects in the correlation structure. Therefore,
it has been of prolonged interest in the literature to provide flexible
and effective ways to model nonstationarity in the spatial effects.
Arguably, due to the local nature of the problem, we might envision
that the correlation structure would be highly dependent on local
characteristics of the domain of study, namely, the latitude, longitude
and altitude of the observation sites, as well as other locally defined
covariate information. In this work, we provide a flexible and
computationally feasible way for allowing the correlation structure of
the underlying processes to depend on local covariate information. We
discuss the properties of the induced covariance functions and methods
to assess its dependence on local covariate information. The proposed
method is used to analyze daily ozone in the southeast United States.
\end{abstract}

%
\begin{keyword}
\kwd{Covariance estimation}
\kwd{nonstationarity}
\kwd{ozone}
\kwd{spatial data analysis}.
\end{keyword}

\end{frontmatter}

\section{Introduction}\label{sintro}

The advance of technology has allowed for the storage and analysis of
complex data sets. In particular, environmental phenomena are usually
observed at fixed locations over a region of interest at several time
points. The literature on modeling spatiotemporal processes has been
experiencing a significant growth in the recent years. The main
objective of this research is to define flexible and realistic
spatiotemporal covariance structures, since predictions for unobserved
locations and future time points, and the corresponding prediction
error variances, are highly dependent on the covariance structure of
the process. An important challenge is to specify a flexible covariance
structure, while retaining model simplicity.

In this paper we are concerned with modeling ozone levels observed in
the southeast USA. We explore
models for ozone which allow the covariance structure to be
nonseparable and nonstationary. Many spatiotemporal models have been
proposed for ambient ozone data for various purposes. \citet
{gms94} and
\citet{mgs98} generate predictions using independent spatial deformation
models for each time period to evaluate deterministic models.
\citet
{carroll97} combine ozone predictions with population data to calculate
exposure indices. \citet{hss04} and \citet{doulezidek} use
a~dynamic
linear model to perform short-term forecasting over a small region,
while \citet{sgh2007} use a dynamic linear model to predict temporal
summaries of ozone and examine meteorologically-adjusted trends over
space. \citet{gillelandnychka2005} seek a method for drawing attainment
boundaries. \citet{mcmillan2005} present a mixture model that allows
heavy ozone production and normal regimes; the probability of each
depends on atmospheric pressure. \citet{downscaler} combine
deterministic model output with observations via a computationally
efficient hierarchical Bayesian approach. \citet{nhm2010} explicitly
model ozone chemistry and transport with additional goals of
decomposition into global background, local creation and regional
transport components, and of long-term prediction under hypothetical
emission controls.

A challenging aspect of modeling ozone is its complex relationship with
meteorology. Tropospheric ozone is a secondary pollutant in that it is
not directly emitted from cars, power plants, etc. Instead, it is
formed from photochemical reactions of precursors nitrogen oxides
(NOx), and volatile organic compounds (VOCs), which are primary
pollutants. The reactions that form ozone are driven by sunlight, so
that ambient concentrations are highest in hot and sunny conditions,
and ozone, NOx and VOCs are transported on the wind, so that emissions
at one site affect ozone at another. It is therefore natural to wonder
whether meteorological variables affect not only the mean
concentration, but also its variance and spatiotemporal correlation. Of
the studies mentioned, \citet{gms94}, \citet{mgs98},
\citet{handh04} and
\citet{nhm2010} model the dependence of the covariance on
covariates in
some form. \citet{gms94} and \citet{mgs98} allow the
spatial covariance
to vary by hour of the day, while \citet{nhm2010} allow it to
vary by
season. \citet{handh04} allow the covariance to vary as a~function of
wind speed and direction, and \citet{nhm2010} model the transport of
ozone using wind speed and direction.

We present a class of spatiotemporal covariance functions that allow
the meteorological covariates to affect the covariance function
[\citet{schmidtguttorpohagan2010}, \citet{schmidt2010}].
This produces a
nonstationary covariance, since the correlation between pairs of points
separated by the same distance may be different depending on local
meteorological conditions.
\citet{sampson92} were among the first to propose a nonstationary
spatial covariance function by making use of a~latent space wherein
stationarity holds. \citet{schmidtohagan2003} proposed a Bayesian
model using
the idea of the latent space where inference is performed under a single
framework. \citet{higdonswallkern1998} use a
moving average convolution approach based on the fact that any Gaussian
process can be represented as a convolution between a kernel and a~white noise
process; nonstationarity results from allowing the kernel to vary
smoothly across locations. \citet{fuentes2002}, instead, assumed that
the spatial
process is a convolution between a fixed kernel and independent Gaussian
processes whose parameters are allowed to vary across locations.
\citet{pacioreckschervish2006} generalize the kernel convolution approach
of \citet{higdonswallkern1998}. On the other hand, \citet{CreHua99},
\citet{Gne02} and \citet{Stein05} present examples of nonseparable
stationary covariance functions
for space--time processes. Although these models provide flexible covariance
structures, they usually have many parameters, which may be challenging
to estimate.

\citet{Cooley2007} capture nonstationarity using
covariates (but not geographic coordinates) to model extreme
precipitation. \citet{schmidtguttorpohagan2010} extended the work of
\citet{schmidtohagan2003} by allowing both geographic coordinates and
covariates to define the axis of the latent space. They also provide a
particular case of the general model which
has a simpler structure but is still able to capture nonstationarity.
\citet{schmidt2010} apply this simpler version of the model in
the case of
multivariate counts observed across the shores of a~lake.

In this paper we provide a more flexible covariance model that allows
not only the distance between covariates, but also the covariate values
themselves to affect the spatial covariance. For example, the spatial
covariance is allowed to be different for a pair of observations with
the same temperature on a cold day than for a pair of observations with
the same temperature on a warm day. Following \citet
{fuentes2002}, we
model the spatial process at location $\bs$, $\mu(\mathbf{s})$, as a
linear combination of stationary processes with different covariances,
%
%
\begin{equation}\label{fuentes}
\mu(\mathbf{s}) = \sum_{j=1}^M w_j(\mathbf{s})\theta_j(\mathbf{s}),
\end{equation}
where $w_j(\mathbf{s})$ are the weights and $\theta_j$ are independent
zero-mean Gaussian processes with covariance $K_j$. \citet{fuentes2002}
models the weights as kernel functions of space centered at predefined
knots $\phi_j$, so that $K_j$ represents the local covariance for sites
near $\phi_j$. In contrast, we specify the weights in terms of spatial
covariates, so that $K_j$ represents the covariance under environmental
conditions described by the covariates.

The paper proceeds as follows. Section \ref{smodel} introduces the
model and Section~\ref{stheory} discusses its
properties. Model-fitting issues and computational details are
discussed in Sections~\ref{spriors} and~\ref{smcmc}, respectively. We
analyze ozone data in Section~\ref{sdata}. We find that the spatial
correlation is stronger on windy days, and that temporal correlation
depends on temperature and cloud cover. Section~\ref{sdisc} concludes.

\section{Covariate-dependent covariance functions}\label{smodel}

Let $y(\bs,t)$ be the observation taken at spatial location $\bs\in
\mathcal{R}^2$ and time $t \in\mathcal{R}$. The response is modeled as
a function of $p$ covariates $\bx(\bs,t)=[x_1(\bs,t),\ldots,x_p(\bs
,t)]^T$, where $x_1(\bs,t)=1$ for the intercept. We assume that
%
%
\begin{equation}\label{ymodel}
y(\bs,t) = \bx(\bs,t)^T\bbeta+ \delta(\bs) + \mu(\bs,t) +
\varepsilon
(\bs,t),
\end{equation}
where\vspace*{1pt} $\bbeta$ is the $p$-vector of regression coefficients, $\delta$
is a Gaussian process to capture the overall spatial trend remaining
after accounting for $\bx(\bs,t)^T\bbeta$ [\citet{Stein97}],
$\mu(\bs
,t)$ is a spatiotemporal effect, and $\varepsilon(\bs,t)\stackrel
{\mathrm{i.i.d.}}{\sim}
\mathrm{N}(0,\sigma^2)$ is pure error.

The spatiotemporal process $\mu$ is taken to be a Gaussian process with
mean zero and covariance that may depend on (perhaps a subset of) the
covariates, $\bx$. As described in Section \ref{sintro}, we model
$\mu
$ as a linear combination of stationary processes,
%
%
\begin{equation}\label{mumodel}
\mu(\bs,t) = \sum_{j=1}^M w_j[\bx(\bs,t)]\theta_j(\bs,t),
\end{equation}
where $\theta_j$ are independent Gaussian processes with mean zero and
covariance~$K_j$ and $w_j[\bx(\bs,t)]$ is the weight on process $j$.
The motivation for this model is that different environmental
conditions, described by the covariates, may favor different covariance
functions. The weight $w_j[\bx(\bs,t)]$ determines the spatiotemporal
locations where the covariance function $K_j$ is the most relevant.

Integrating over the latent processes $\theta_j$, the covariance becomes
%
%
\begin{equation}\label{cov1}\quad
\operatorname{Cov}[\mu(\bs,t),\mu(\bs',t')|\bx] =
\sum_{j=1}^M w_j[\bx(\bs,t)]w_j[\bx(\bs',t')]K_j(\bs-\bs',t-t').
\end{equation}
With $M=1$, only the variance of the process depends on the covariates,
and the correlation, $K_1(\bs-\bs',t-t')/K_1(0,0)$, is stationary. With
$M>1$, both the variance and the correlation depend on the
covariates.\vadjust{\goodbreak}

\begin{figure}

\includegraphics{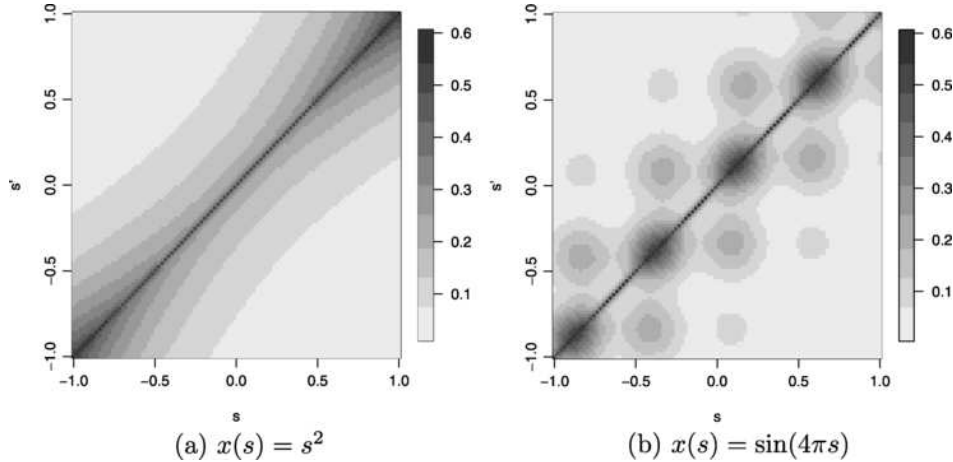}

\caption{Covariance functions for a one-dimensional spatial process
with $M=2$, $\operatorname{logit}(w_2(s))= x(s)$, $w_1(s)=1-w_2(s)$, $K_1(s-s') =
\exp(-|s-s'|/0.02)$, and $K_2(s-s') =
\exp(-|s-s'|/0.50)$.}\label{fexample}
\end{figure}

As an illustration of the flexible spatial patterns allowed by our
specification, Figure \ref{fexample} plots the spatial covariance for
two simple examples. In both cases we assume a one-dimensional spatial
grid with $s\in\mathcal{R}$, a single covariate~$x(s)$, and that the
spatial correlation is high in areas with large $x(s)$. Both examples
have $M=2$, $\operatorname{logit}(w_2(s))=x(s)$, $w_1(s)=1-w_2(s)$, $K_1(s-s') = \exp
(-|s-s'|/0.02)$, and $K_2(s-s') = \exp(-|s-s'|/0.50)$. Figure~\ref
{fexample} shows the covariance for $x(s)=s^2$ and $x(s)=\sin(4\pi
s)$. For the quadratic covariate, the second term has higher spatial
correlation and the weight on the second process is high for locations
with large~$x(s)$, therefore, the spatial correlation is stronger for
$s$ near $-1$ and 1 where $x(s)$ is high. The spatial covariance is not a
monotonic function of spatial distance for the periodic covariate. This
may be reasonable if, say, $x(s)$ is elevation and a site with high
elevation shares more common features with other high-elevation sites
than nearby low-elevation sites.

%

There is confounding in (\ref{cov1}) between the scale of the weights
$w_j$ and covariances $K_j$, since multiplying the weights by the
constant $c>0$ and dividing the standard deviation of $K_j$ by $c$
gives the same covariance. Therefore, for identification purposes we
restrict the squared weights for each observation to sum to one, $\sum
_{j=1}^M w_j[\bx(\bs,t)]^2 = 1$. Also, allowing the weights to be
negative would result in a negative spatiotemporal covariance if
$w_j[\bx(\bs,t)]>0$ and $w_j[\bx(\bs',t')]<0$. In some situations this
may be desirable, however, we elect to restrict $w_j[\bx(\bs,t)]>0$ to
ensure a positive spatiotemporal covariance. Section \ref{spriors}
discusses weight selection in further detail.

An important consequence of the covariance construction in (\ref
{mumodel}) is that values of the process at two sites are uncorrelated
unless at least one of the $M$ weight functions is positive at both
sites. Therefore, unlike other nonstationary covariance models [e.g.,
\citet{sampson92} and \citet{higdonswallkern1998}], it may be
difficult to separate strength of dependence from severity of
nonstationarity. For example, if $M=2$, $w_1(\bs_1)=w_2(\bs_2)=1$, and
$w_2(\bs_1)=w_1(\bs_2)=0$, then not only is the covariance near~$\bs_1$
different than the covariance near $\bs_2$, but $\mu(\bs_1)$ and
$\mu
(\bs_2)$ are necessarily uncorrelated. If this is deemed undesirable
for a particular application, one alternative would be to allow for
dependence between the latent $\theta_j$ using a~multivariate spatial
model. Another option would be to use only covariates in the weights
that have larger spatial range (perhaps pre-smoothed covariates) than
the latent~$\theta_j$ processes, in which case this scenario is less
likely. Section~\ref{ssmooth} provides further discussion about the
relative roles of the spatial range of the latent and covariate processes.

This covariance model has interesting connections with other commonly
used spatial models. For example, if we consider purely spatial data,
as mentioned in Section \ref{sintro}, taking the weights to be kernel
functions of the spatial location alone, that is, $w_j[\bx(\bs
)]=w_j(\bs
)$, gives the nonstationary spatial model of \citet{fuentes2002}. By
modeling the weights as functions of the covariates, it may be possible
to explain nonstationarity with fewer terms, giving a~more concise and
interpretable model. Also, with $M=p$ and $w_j[\bx(\bs)]=x_j(\bs)$ for
$j=1,\ldots,p$, we obtain the spatially-varying coefficient model of
\citet
{gelfand2003}. In this model $\theta_j(\bs)$ represents the effect of
the $j$th covariate at location $\bs$. The motivation for the
spatially-varying coefficients model is to study local effects of
covariates on the mean response. In contrast, our objective is to model
the covariance. For example, in a situation with $p=20$ covariates it
may be sufficient to describe the spatial covariance using $M=2$
stationary processes where conditions that favor the two covariance
functions are described by weights $w_1$ and $w_2$ that depend on all
$p$ covariates. Therefore, to provide an adequate description of the
covariance, we assume the weights are random functions of unknown
parameters that describe environmental conditions (see Section \ref
{spriors}) rather than taking the weights to be the covariates
themselves. Finally, setting the weights $w_j$ to be constant in time
and the latent processes $\theta_j$ to be constant over space gives the
spatial dynamic factor model of \citet{Lopes2008}. Our model differs
from this approach since our weights (loadings) are functions of
spatial covariates rather than purely stochastic spatial processes.

\section{Properties of the covariance model}\label{stheory}

In this section we discuss some properties of the proposed model in
(\ref{mumodel}) and the spatiotemporal covariance function. For
example, it is clear that even if the individual covariances~$K_j$ are
separable, stationary and isotropic, the resulting covariance~(\ref
{cov1}) is in general nonseparable, nonstationary and anisotropic.
Below we discuss other properties of the covariance model.

\subsection{Monotonicity of the spatial covariance function}\label{smono}

As shown in Figure \ref{fexample}, the covariance function can be a
nonmonotonic function of spatial distance, even if the underlying
covariances $K_j$ are decreasing. Intuitively, this occurs only if the
spatial range of the covariates is small relative to the spatial range
of the covariance functions $K_j$. More formally, assuming $\bs\in
\mathcal{R}$ and the $w_j$ and each component of $\bx$ are
differentiable, then for any $h>0$
%
%
\begin{eqnarray}\label{CovPrime}
&&\frac{\partial\operatorname{Cov}(\mu(\bs),\mu(\bs+h)|\bx)}{\partial h}
\nonumber\\[-8pt]\\[-8pt]
&&\qquad= \sum
_{j=1}^M w_j(\bx[s])w_j(\bx[s+h])K_j(h)\biggl[\frac{w_j'(\bx
[s+h])}{w_j(\bx[s+h])} + \frac{K_j'(h)}{K_j(h)}\biggr],\nonumber
\end{eqnarray}
both\vspace*{1pt} $w_j'$ and $K_j'$ are derivatives with respect to $h$. Therefore,
if the weights $w_j(\bx[s])$ and covariance $K_j(h)$ are positive, a
sufficient but not necessary condition for a monotonic covariance is
that $w_j'(\bx[s+h])/w_j(\bx[s+h]) + K_j'(h)/ K_j(h)<0$ for all $j$. The
ratios $w_j'(\bx[s])/w_j(\bx[s])$ and $-K_j'(h)/K_j(h)$ can be
interpreted as the elasticity of the weight function (which depends on
both the weight function itself and the derivative of the covariate
process) and covariance function, respectively. This condition makes
the initial statement more precise, in that (\ref{CovPrime}) is
negative if the elasticity of the weight function is less than the
elasticity of the spatial covariance.

In the special case of a powered-exponential covariance model $K_j(\bs
,\bs+h) = \tau_j^2\exp(-\rho_jh^{\kappa_j})$ and exponential weights
$w_j(\bx) = \exp(\bx^T\balpha_j)$, where $\balpha_j$ is a vector of
coefficients, (\ref{CovPrime}) becomes
%
%
\begin{eqnarray}\label{CovPrime2}
&&
\frac{\partial\operatorname{Cov}(\mu(\bs),\mu(\bs+h)|\bx)}{\partial h}
\nonumber\\[-8pt]\\[-8pt]
&&\qquad= \sum_{j=1}^M
w_j(\bx[s])w_j(\bx[s+h])K_j(h)[\Delta_x(\bs+h)^T\balpha
_j-\kappa
_j\rho_jh^{\kappa_{j}-1}],\nonumber
\end{eqnarray}
where $\Delta_x(\bs+h)$ denotes the vector of derivatives of $x(\bs+h)$
with respect to~$h$.
The covariance is decreasing in~$h$ if $\Delta_x(\bs+h)\balpha
_j<\kappa
_j\rho_jh^{\kappa_{j}-1}$ for all~$j$ and $h$.
This shows that it is possible to allow the spatial covariance to
depend on covariates but retain monotonicity by restricting the
parameters~$\balpha_j$, $\kappa_j$ and $\rho_j$ based on bounds on the
covariate process derivatives.

\subsection{Smoothness properties of the spatial process}\label{ssmooth}

The smoothness properties of a Gaussian process are often quantified in
terms of the mean squared continuity of its derivatives. For many
spatial processes, including the nonstationary model of \citet
{fuentes2002}, the smoothness of their process realizations is well
studied [see \citet{BG2003}, \citet{BGS2003}]. However, our model
postulates a more general dependence of the covariance on spatial
covariates. Hence, in this section we explore the effect of that
dependence on the smoothness properties of the realizations. For
notational convenience, we assume a one-dimensional spatial process
with $s\in\mathcal{R}$; the results naturally extend to more general
direction derivatives by taking $s=\bu^T\bs$ for any unit vector $\bu$.
We start by assuming the covariates $\bx$ are fixed; this assumption
will be later relaxed.

Following the arguments of \citet{BG2003}, we say that the $k$th
derivative (with respect to $s$) of the process $\mu$ (if it exists) is
mean square continuous at $s$ if
%
%
\begin{equation}\label{MScont}
\lim_{h\rightarrow0}\mathrm{E}\bigl[\mu^{(k)}(s)-\mu^{(k)}(s+h)|\bx
\bigr]^2=0.
\end{equation}
For $k=0$, we can substitute (\ref{mumodel}) in (\ref{MScont}) and get
%
%
\begin{eqnarray}\label{MScont2}
&&
\lim_{h\rightarrow0}\mathrm{E}[\mu(s)-\mu(s+h)|\bx]^2\nonumber\\
&&\qquad=
\sum_{j=1}^M\lim_{h\rightarrow0}K_j(0)\bigl(w_j[\bx(s+h)]-w_j[\bx
(s)]\bigr)^2 \\
&&\qquad\quad{}+ \sum_{j=1}^M\lim_{h\rightarrow0}2w_j[\bx(s)]w_j[\bx
(s+h)]\bigl(K_j(0)-K_j(h)\bigr),\nonumber
\end{eqnarray}
which shows that $\mu$ is mean square continuous if each latent process
is mean square continuous [$\lim_{h\rightarrow0}K_j(h)=K_j(0)$] and
the weights are smooth enough to satisfy $\lim_{h\rightarrow
0}(w_j[\bx
(s+h)]-w_j[\bx(s)])^2=0$ for all $j$, for example, they are continuous
functions of the continuous spatial covariates.

In some settings, it may be reasonable to consider $\bx$ to be a random
process. We extend the discussion of \citet{BG2003} to the case
when the
weights are functions of stochastic covariates. In this case, to study
the smoothness of $\mu$ requires considering variability in both the
latent $\theta_j$ as well as the covariates~$\bx$. The covariates enter
the covariance model only through the stochastic weights $W_j(s) =
w_j[\bx(s)]$. Taking the expectation with respect to both $\theta_j$
and $W_j(s)$ gives
%
%
\begin{eqnarray}\label{MScont2}
&&
\lim_{h\rightarrow0}\mathrm{E}[\mu(s)-\mu(s+h)]^2\nonumber\\
&&\qquad=
\sum
_{j=1}^M\lim_{h\rightarrow0}K_j(0)E_{W_j}[W_j(s)-W_j(s+h)]^2\\
&&\qquad\quad{}+2\sum_{j=1}^M\lim_{h\rightarrow0}\bigl(K_j(0)-K_j(h)\bigr)
\mathrm{E}_{W_j}[W_j(s)W_j(s+h)].\nonumber
\end{eqnarray}
Therefore, under stochastic covariates, the process $\mu$ is mean
square continuous if and only if the latent processes $\theta_j$ and
the weight processes $W_j$ are both mean square continuous. It is well
known from probability theory that the weight function $W_j$ is mean
square continuous, for example, if it is bounded and the covariate
processes are almost surely continuous. Mean square continuity also
follows when $w_j$ is Lipschitz continuous of order 1 and the covariate
processes are mean square continuous. For example, the logistic weights
$w_j(\bx) = \exp(\bx^T\balpha_j)/[1+\exp(\bx^T\balpha_j)]$ are both
bounded and Lipschitz continuous of order 1, whereas exponential
weights $w_j(\bx) = \exp(\bx^T\balpha_j)$ are not.\looseness=-1

These results naturally extend from mean squared continuity to mean
square differentiability, and higher order derivatives. Since $\mu(s)$
is the sum of stochastic processes $Z_{j}(s)=W_{j}(s)\theta_{j}(s)$,
then $\mu^{(k)}(s)=\sum_{j=1}^{M}Z_{j}^{(k)}(s)$. In particular, for
\mbox{$k=1$} the derivative process at $s$
is
%
%
\begin{equation}
\mu^{(1)}(s)=\sum_{j=1}^{M}\theta_{j}^{(1)}(s)W_{j}(s)+\theta
_{j}(s)W^{(1)}_{j}(s).
\end{equation}
So the process $\mu$ is mean square differentiable if both $W_{j}(s)$
and $\theta_{j}(s)$ are mean
square differentiable. Conditions analogous to those outlined above for
mean square
continuity will assure that the weights are mean square differentiable.
More precisely, if the covariate processes $x_1(s), \ldots, x_p(s)$
are mean square differentiable and the function $w_j(\cdot)$ is
Lipschitz continuous of order 1, then the resulting process $W_j(s)$ is mean
square differentiable, and so is $\mu(s)$.

One could go further to study sample path properties and almost sure
continuity of the induced spatial process, although the required proofs
are generally more difficult than the proofs of mean square properties.
If both the weight functions and latent processes are almost surely
continuous, then the induced spatial process is also almost surely
continuous. \citet{adler1981} and \citet{kent1989} provide
conditions to
verify almost sure continuity for spatial fields.

\subsection{Span of the covariance function}

The covariance in (\ref{cov1}) is quite flexible. For example, consider
partitioning the covariate space in $N$ subsets $\mathcal{A}_1,
\ldots, \mathcal{A}_N$ and
\[
w_{j}[\mathbf{x}(\mathbf{s},t)]=\sum_{i=1}^N a_{ji}I\bigl(\mathbf
{x}(\mathbf{s},t) \in A_i\bigr).
\]
When $\mathbf{x}(\mathbf{s},t) \in A_i$ and $\mathbf{x}(\mathbf
{s}',t') \in A_{k}$, the
covariance becomes
\[
\operatorname{Cov}(\mu(\mathbf{s},t),\mu( \mathbf{s}',t')|\mathbf{x})=\sum
_{j=1}^M a_{ji} a_{jk}
K_{j}(\mathbf{s}-\mathbf{s}', t-t').
\]
Hence, each covariance $ K_{j}(\mathbf{s}-\mathbf{s}', t-t')$
contributes to the
mixture differently according to the levels of the covariates. Setting
some of the weights $a_{ij}=0$ allows $K_j$ to contribute only to the
covariance of terms with specific combinations of covariates, for
example, both observations have low wind speed and high cloud cover.
Also, by setting some of the $a_{ij}<0$, it is possible to specify
negative correlation for observations with different levels of the
covariate. By increasing $M$ and $N$, this argument shows how the
covariate-dependent weights can be used to describe quite general
spatiotemporal behavior depending on the covariates.

\section{Priors and model-fitting}\label{spriors}

In this section we describe a convenient specification of the model.
For notational convenience, we assume that at each time point
observations are taken at spatial locations $\bs_1,\ldots,\bs_N\in
\mathcal
{R}^2$ and that $t\in\{1,2,\ldots\}$. The overall spatial trend
$\delta$
is a Gaussian process with mean zero and spatial covariance $K_0^s$. We
assume that $\delta$'s covariance is stationary, although one could
allow $\delta$'s covariance to be nonstationary as well. We assume an
autoregressive spatiotemporal model for the latent processes $\theta_j$,
%
%
\begin{equation}\label{DLM}
\theta_j(\bs,t) = \gamma_j\theta_j(\bs,t-1)+e_j(s,t),
\end{equation}
where $\gamma_j\in(0,1)$ controls the temporal correlation and the
$\mathbf e_{jt}=[e_j(\bs_1,t),\ldots,\allowbreak e_j(\bs_N,t)]$ are independent
(over $j$
and $t$) spatial processes with mean zero and spatial covariance
$K_j^s$. We use exponential covariance functions for $K_j^s$,
$j=0,\ldots,M$. We note that although this is a relatively simple
specification for the temporal component for each latent process,
complex temporal covariance structures can emerge from this mixture
model. The covariance between subsequent observations at a site is a
mixture of $M$ autoregressive covariances that varies with space and
time according to the covariates. This approach could be very useful
for modeling hourly ozone which is generally low and steady at night,
and high and volatile in the day, which could be fit by including hour
of the day as a covariate in the weights.


As mentioned in Section \ref{smodel}, there is confounding in (\ref
{cov1}) between the scale of the weights $w_j$ and covariances $K_j$.
Therefore, for identification purposes we restrict the squared weights
for each observation to sum\vspace*{1.5pt} to one, $\sum_{j=1}^M w_j[\bx(\bs,t)]^2 =
1$. Although there are other possibilities, we assume the weights have
the multinomial logistic form
%
%
\begin{equation}\label{weights}
w_j[\bx(\bs,t)]^2 = \frac{\exp{(\bx(\bs,t)^T\balpha
_j)}}
{\sum_{l=1}^M\exp{(\bx(\bs,t)^T\balpha_l)}},
\end{equation}
where $\balpha_1,\ldots,\balpha_M$ are vectors of regression
coefficients that control the effects of the covariates on the
covariance. For these weights setting $M=1$ gives $w_1[\bx(\bs,t)]=1$
and the model is stationary with covariance $K_1$. The choice of
logistic weights also ensures mean square continuity of the process
realizations, as outlined in Section \ref{stheory}. For identification
purposes, we fix $\balpha_1=0$, as is customary in logistic regression.

The priors for the hyperparameters are uninformative. We use
$\mathrm{N}(0,10^2)$ priors for the elements of $\bbeta$ and $\balpha_{j}$. The
covariance\vspace*{-1pt} parameters have priors
$\sigma^{-2},\tau_j^{-2} \stackrel {\mathrm{i.i.d.}}{\sim}
\operatorname{Gamma}(0.1,0.1)$, and $\gamma_j\sim
\operatorname{Unif}(0,1)$. Also,\vspace*{1pt} we take $K^s_j(\|h_s\|) =
\exp(-\|h_s\|/\rho_j)$, where $h_s$ is the distance between points
after a Mercator projection, scaled to correspond roughly to distance
in kilometers, and $\rho_j\sim \operatorname{Unif}(0,2\mbox{,}000)$.

The covariance and the effect of an individual covariate on the
covariance in (\ref{cov1}) are rather obscure. This is due to the
label-switching problem, that is, the labels of the processes are
arbitrary: for example, $\theta_1$ may correspond to a high variance
process for some MCMC iterations and to a small variance process for
others, making inference on individual parameters difficult. One remedy
for the label-switching problem is to introduce constraints, perhaps,
$\operatorname{Var}(\theta_1)<\cdots< \operatorname{Var}(\theta_M)$. However, ordering
constraints on
complex functions such as spatiotemporal covariance functions is not
straightforward. Therefore, rather than summarizing the individual
parameters in the model, we summarize the entire covariance function
for different combinations of covariates. A simple way to summarize the
effect of the $k$th covariate is in terms of the posterior of the
ratio of the covariance of two observations with $x_k=2$ (standard
deviation units above the mean) compared to the covariance of two
observations with $x_k=0$, assuming all other covariates are fixed at
zero (their mean). That is,
%
%
\begin{equation}\label{Delta}
\Delta_k(h_s,h_t) = \frac{
\sum_{j=1}^M (({\exp(\alpha_{j1}+\alpha_{jk})})/({\sum
_{l=1}^M\exp
(\alpha_{l1}+\alpha_{lk})}))K_j(h_s,h_t)}
{\sum_{j=1}^M (({\exp(\alpha_{j1})})/({\sum_{l=1}^M\exp(\alpha
_{l1})}))K_j(h_s,h_t)},\hspace*{-35pt}
\end{equation}
where $\alpha_{jk}$ is the $k$th element of $\balpha_j$ and
$K_j(h_s,h_t)=K_j^s(\|h_s\|)\gamma_j^{|h_t|}$. We also\vspace*{-1pt} inspect the
ratio of correlations ${\tilde\Delta}_k(h_s,h_t) = \Delta
_k(h_s,h_t)/\Delta_k(0,0)$. We consider a covariate to have a
significant effect on the variance if the posterior interval for
$\Delta
_k(0,0)$ excludes one. Similarly, we consider a covariate\vspace*{2pt} to have a
significant effect on the spatial (temporal) correlation if the
posterior interval for ${\tilde\Delta}_k(h_s,0)$ [${\tilde\Delta
}_k(0,h_t)$] excludes one.

Finally, we discuss how to select the number of terms, $M$. One
approach would be to model $M$ as unknown and average over model space
using reversible jump MCMC. Lopes, Salazar and Gamerman
(\citeyear{Lopes2008}), Salazar, Lopes
and Gamerman (\citeyear{Salazar2009}) use
reversible jump MCMC to select the number of factors in a latent
spatial factor model. However, this approach is likely to pose
computational challenges for large spatiotemporal data sets. Therefore,
we select the number of terms using cross-validation
and assume $M$ is fixed in the final analysis. For cross-validation, we
randomly (across space and time) split the data into training
($n=63$,881) and testing ($N=3$,367) sets. We fit the model on the training
data and compute the posterior predictive distribution for each test
set observation. We then compute\vadjust{\goodbreak} the mean squared error $\mathrm{MSE}=\sum
_i(Y_i-{\bar Y}_i)^2/N$ and mean absolute deviation $\mathrm{MAD}=\sum
_i(Y_i-{\tilde Y}_i)^2/N$, where the sum is over the $N$ test set
observations,~${\bar Y}_i$ is the posterior mean, and~${\tilde Y}_i$ is
the posterior median. We also compute the mean (over the test set
observations) of the posterior predictive variances (``AVE VAR''), the
median of the posterior predictive standard deviations (``MED SD'') and
the coverage probability of 95\% prediction intervals.


\section{Computational details}\label{smcmc}

We implement the model in R (\href{http://www.r-project.org/}{http://www.}
\href{http://www.r-project.org/}{r-project.org/}). Though
implementation in WinBUGS (\href{http://www.mrc-bsu.cam.ac.uk/bugs/}{http://www.mrc-}
\href{http://www.mrc-bsu.cam.ac.uk/bugs/}{bsu.cam.ac.uk/bugs/})
would also be straightforward, run times might be long for large data
sets. We update $\btheta_{jt}=[\theta_j(\bs_1,t),\ldots,\theta
_j(\bs
_N,t)]$, $\bbeta$, $\sigma^2$ and~$\gamma_j$, which have conjugate full
conditionals, via Gibbs sampling, and we update $\alpha_{jk}$, $\rho_j$
and $\nu_j$ using Metropolis--Hastings sampling with a Gaussian
candidate distribution tuned to give acceptance probability around 0.4.

Sampling using the dynamic spatial model in (\ref{DLM}) allows us to
update the $\btheta_{jt}$ as a block and avoid inverting large
matrices. The alternative of sampling after marginalizing out the
latent $\btheta_{jt}$ would require computing the entire spatiotemporal
covariance with elements given by (\ref{cov1}), which would likely give
better mixing for small to moderate data sets. For our large data set,
however, matrix computations of this size are not feasible.

We monitor convergence with trace and autocorrelation plots for several
representative parameters. Monitoring convergence is challenging for
this model since the labels of the latent terms may switch during MCMC
sampling: exchanging $\balpha_1$, $\rho_1$, $\nu_1$ and $\gamma_1$, for
example, with $\balpha_2$, $\rho_2$, $\nu_2$ and~$\gamma_2$, does not
affect the covariance in (\ref{cov1}). Rather than monitoring
convergence for these parameters individually, we therefore monitor
convergence of the covariance (\ref{cov1}) at several lags and of the
spatiotemporal effect $\mu(\bs,t)$ for several spatiotemporal
locations. For the application in Section \ref{sdata} we generate
20,000 samples, discarding the first 10,000 as burn-in. For the
exponential covariances considered here this appears to be sufficient;
however, for smoother processes, such as those induced by the squared
exponential covariance, 20,000 iterations may not be sufficient.

\section{Application to southeastern US daily ozone}\label{sdata}

\begin{figure}

\includegraphics{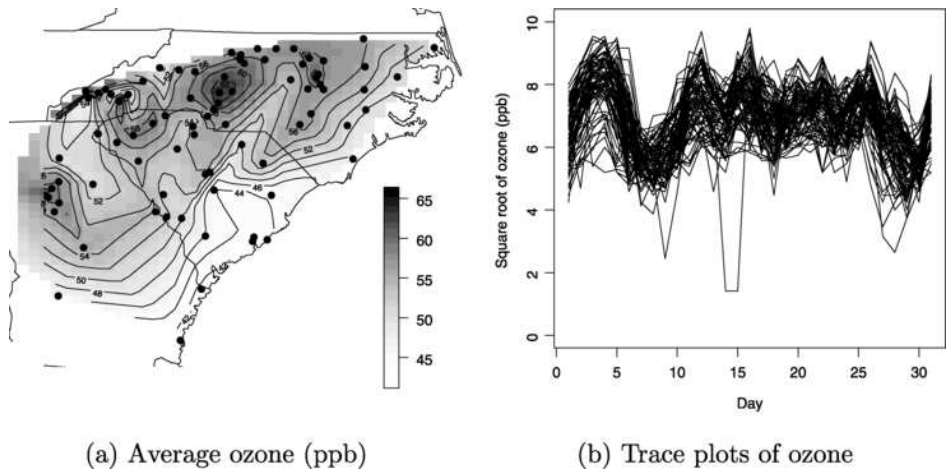}

\caption{Plots of square root ozone (pbb). Panel \textup{(a)} plots
the average for each station (the stations are marked with points) and
panel \textup{(b)} gives trace plots for each station in August 2005 (Day 1 is
August 1, 2005).}\label{fdataplots}
\end{figure}

To illustrate our spatiotemporal covariance model, we analyze ozone in
the southeast US. The primary National Ambient Air Quality Standard
(NAAQS) for ozone requires the three-year average of the annual
fourth-highest daily maximum 8-hour daily average concentration to fall
beneath 75 parts per billion (ppb) [\citet{cfr}, pages 16436--16514].
Our response variable is thus the square root---to ensure
Gaussianity---of the daily ``8-hour ozone'' metric. We focus on the $89$
sites in North Carolina, South Carolina and Georgia shown in Figure
\ref
{fdataplots}. This geographically heterogeneous region transitions
from the flat,\vadjust{\goodbreak} low-altitude coastal plains in the east, to the gentle,
rolling hills of the piedmont, to mountains in the northwest, with a
handful of urban islands buffered by suburbs that give way to rural
tracts. Since summertime ozone concentrations are highest, and
therefore most relevant for attainment determination, we extract daily
8-hour ozone concentrations, longitude, latitude, elevation and site
classification (urban, suburban or rural) for June--August, 1997--2005
($6444/73$,$692=8.7$\% missing) from the US EPA Air Quality System (AQS)
database, available via the Air Explorer web tool
(\href{http://www.epa.gov/airexplorer/index.htm}{http://}
\href{http://www.epa.gov/airexplorer/index.htm}{www.epa.gov/airexplorer/index.htm}).

%

We obtain daily average temperature and daily maximum wind speed from
the National Climatic Data Center (NCDC) Global Summary of the Day and
daily average cloud cover from the NCDC National Solar Radiation
database. Since meteorological and ozone data are not observed at the
same locations, for each day we predict each meteorological variable at
ozone observation sites using spatial Kriging implemented in SAS V9.1
Proc MIXED with an exponential covariance function. Though \citet
{Li2009} show that ignoring uncertainty when using spatial predictions
of covariates is not without consequence, accounting for that
uncertainty is nontrivial. Since our current focus is the development
of the covariate-dependent covariance model, we treat these predictions
as fixed.

Covariates $\bx(\bs,t)$ in the mean trend include the continuous
variables temperature, wind speed, cloud cover, elevation, longitude,
latitude and a linear trend in year, each standardized to have mean
zero and variance one, and we include two indicator variables
identifying a station as urban or rural, leaving suburban\vadjust{\goodbreak} as the
baseline. We have no detailed land-use covariates as in \citet
{Paciorek09}, however, which would likely improve fine-scale
prediction. We consider all two-way interactions between the three
meteorological variables and quadratic effects of the meteorological
variables. The covariance is modeled as a function of only the main
effects of these covariates.\looseness=-1

\subsection{Empirical variogram analysis}\label{svariogram}

We begin studying the data by analyzing the spatial variogram, defined
as $\gamma(h) = E([r(\bs,t)-r(\bs+ h\bu,t)]^2)$, where
$r(\bs,t)$ is the residual after accounting for the mean trend and
$\bu
$ is a~unit vector. Though there is evidence that regression
coefficient estimation can be affected by disregarding spatial
correlation [\citet{Reich06}, \citet{Wakefield07},
\citet{Paciorek10}],
for simplicity we use ordinary least squares, pooled over all
observations, to estimate the mean trend. We estimate the variogram as
the mean squared difference between all pairs of observations in a bin
$D_h$, that is,
%
%
\begin{equation}\label{variog1}
{\hat\gamma}(h) = \frac{1}{|D_h|}\sum_t\sum_{ (\bs,\bs') \in
D_h}[r(\bs,t)-r(\bs',t)]^2,
\end{equation}
where $D_h$ is the set of pairs of points on the same day with $\|\bs
-\bs'\| \in(h-\epsilon,h+\epsilon)$ and $|D_h|$ is the cardinality
of $D_h$.

%
\begin{figure}

\includegraphics{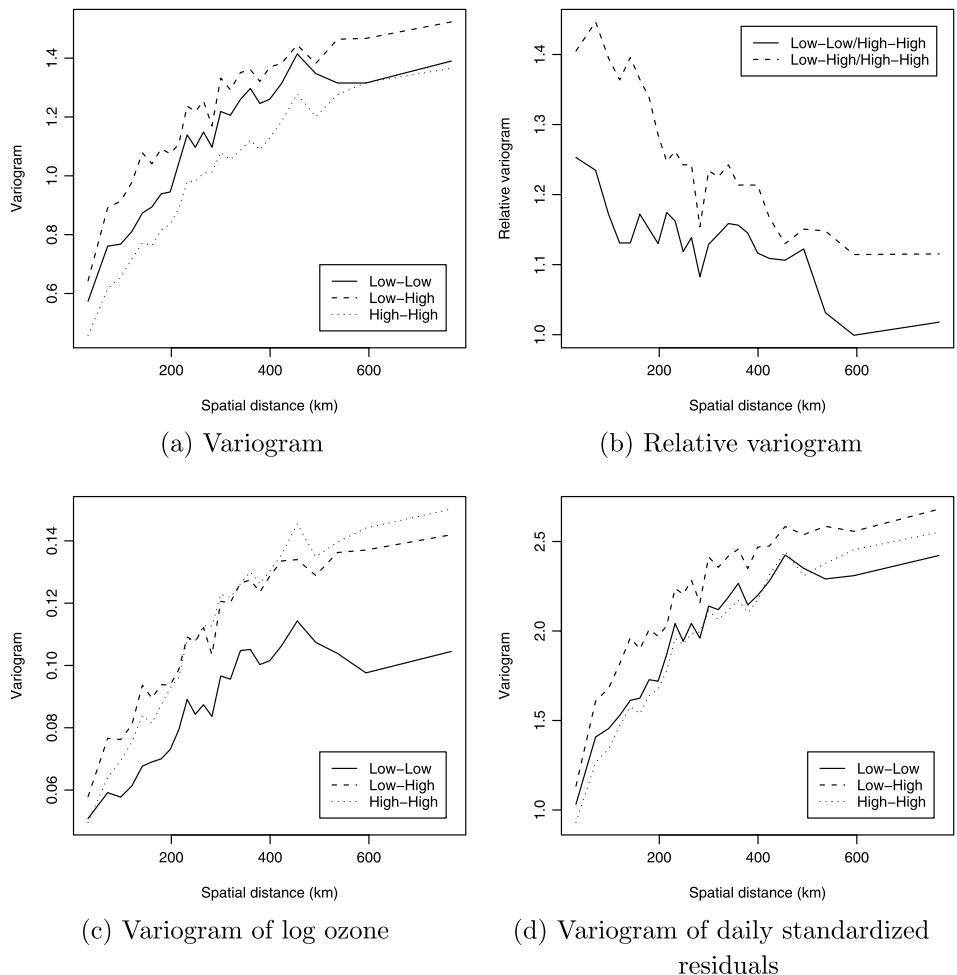}%
\vspace*{3pt}
\caption{Sample variograms for the ozone data by cloud
cover. The data are plotted separately for pairs of observations with
both (``Low--Low''), one (``Low--High'') and neither (``High--High'')
members of the pair with cloud cover below the median cloud cover.
Panel~\textup{(b)} plots the ratio of curves in \textup{(a)}, panel
\textup{(c)} uses
log-transformed, rather than square-root-transformed data, and panel
\textup{(d)} standardizes the residuals by the daily sample standard
deviation.}\label{fvariog1}
\vspace*{6pt}
\end{figure}

%
\begin{figure}

\includegraphics{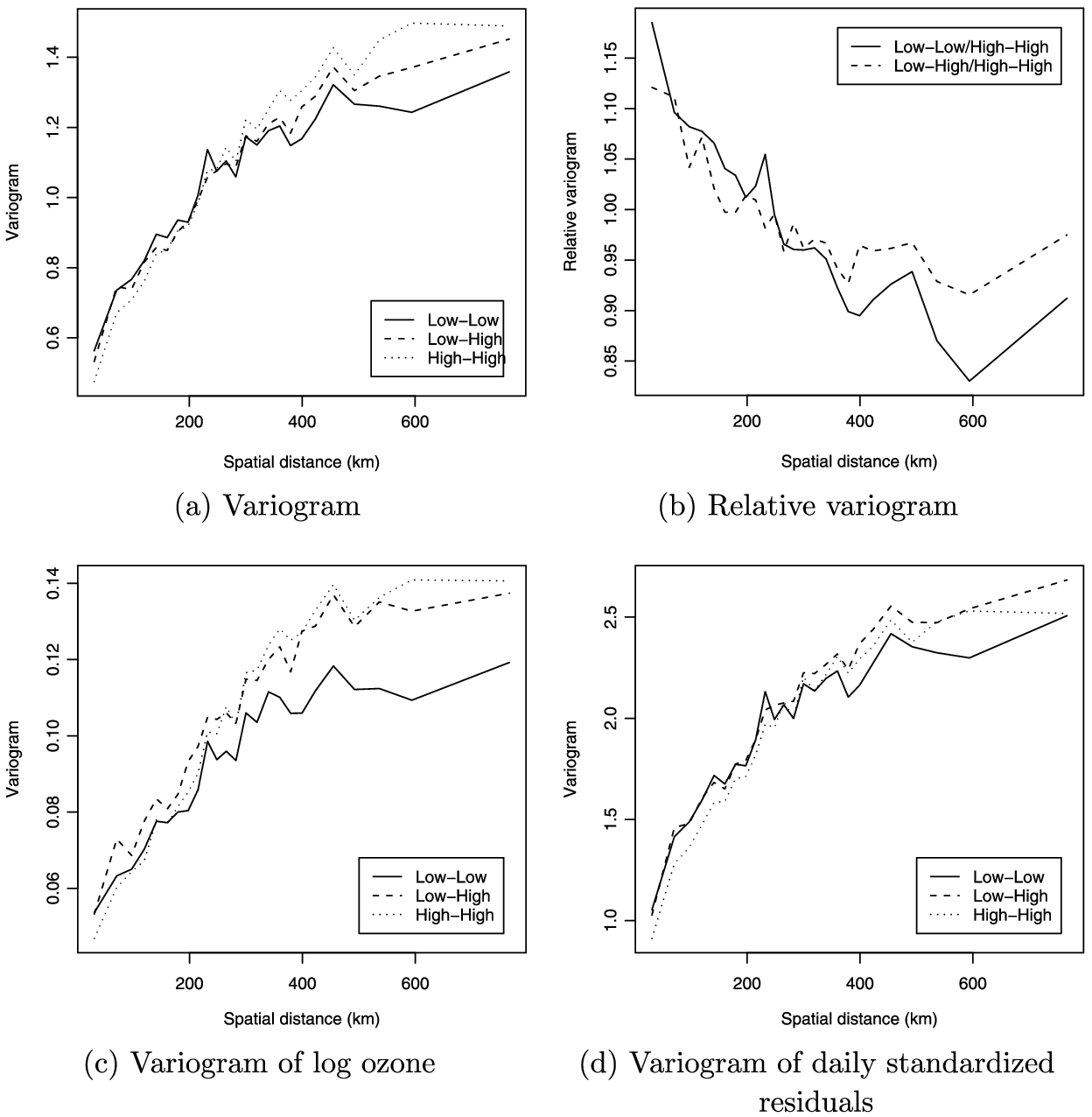}

\caption{Sample variograms for the ozone data by wind speed.
The data are plotted separately for pairs of observations with both
(``Low--Low''), one (``Low--High'') and neither (``High--High'') members
of the pair with wind speed below the median wind speed. Panel~\textup{(b)}
plots the ratio of curves in \textup{(a)}, panel \textup{(c)} uses log-transformed,
rather than square-root-transformed data, and panel \textup{(d)} standardizes
the residuals by the daily sample standard deviation.}\label{fvariog2}
\end{figure}

To explore the effects of each of the covariates on the spatial
covariance, we compute individual variograms for three categories of
site pairings. In the ``low--low'' category, both sites have values of
the covariate below the sample median for the covariate; in the
``high--high'' category, both have values above the median; and in the
``low--high'' category, one has a value below, and the other above, the
median. Such variograms for cloud cover and wind speed are given in
Figures \ref{fvariog1} and \ref{fvariog2}, respectively.

In Figure \ref{fvariog1}(a), the variogram is lowest for pairs of
observations for which both sites have high cloud cover, higher when
both sites have low cloud cover, and highest when one site has low and
the other high cloud cover. Solar radiation is required to turn NO$_2$
into ozone or to create VOC's that turn NO into NO$_2$. Therefore,
under high cloud cover conditions, ozone levels would be expected to
drop close to background levels (a long-term equilibrium that would
exist in the absence of local emissions), which would be homogeneous
over a region of this size. Two sites for which cloud cover is low
would be expected to be less similar to each other than would two sites
that both have high cloud cover because the production of ozone via
solar radiation is now dependent on the spatially-varying precursors.
For example, areas very close to major sources of NOx (power plants and
urban centers on workdays) would have low ozone due to NOx scavenging,
and moving downwind from these sources, ozone would increase and then
decrease. Finally, based on the explanation above, it is clear that if
one site has high cloud cover, so that ozone production is minimal, and
the other has low cloud cover, so that ozone production is rampant,
they would have very dissimilar ozone values, so that the variogram
would be highest for the low--high category.\vadjust{\goodbreak}


Wind speed does not generally affect the chemical reactions that create
or destroy ozone, but it does transport ozone and its precursors. One
would expect that within smaller subregions with higher wind speeds,
distance is effectively shortened so that spatial correlation would be
higher, and two sites in the ``high--high'' category would have lower
variogram, followed by those with ``low--high,'' then ``low--low,'' as we
see in Figures \ref{fvariog2}(a) and \ref{fvariog2}(b) for spatial lags
below 250 km. The ordering of the categories is reversed for larger
spatial lags, where transport is less relevant.


In addition to computing these variograms for our square root ozone
response, we compute the variograms using residuals from a regression
on the log, rather than square root, of ozone, and the variogram of
standardized residuals, that is, $r^*(\bs,t)=r(\bs,t)/s_t$, where $s_t$
is the sample standard deviation of the residuals for day $t$. The
variograms are affected more by the log transformation than by
standardizing. The same general patterns remain after standardizing,
but new ones emerge after a log transformation. For example, the
ordering of the variograms for cloudy and sunny days switches after a
log transformation in Figure \ref{fvariog1}. The patterns of the
log-transformed responses also indicate covariate-dependent covariance,
so it appears that the transformation is important, but does not
resolve nonstationarity.

\subsection{Results}

We fit five versions of the model, with the number of mixture
components varying from $M = 1$ to $5$. We withheld 5\% of the
observations (3,687 observations), selected randomly across space and
time. Table \ref{tmse} compares for predictions of square root ozone
for this validation set. For all models, the prediction intervals have
coverage greater than 0.95. The five-component model minimizes all
measures of prediction error and variance. The ratio of mean squared
error for the five-component model to that of the stationary
one-component model is $0.179/0.189 = 0.947$, and the corresponding ratio
of average prediction variances is $0.167/0.183 = 0.913$. The
nonstationary covariance thus gives a modest improvement in prediction
accuracy and uncertainty quantification. We also tried higher values of
$M$ and found slight improvements in prediction, but elected to proceed
with $M=5$ for model simplicity.

%
\begin{table}
\caption{Validation set results. The summaries are mean
squared error (MSE), median absolute deviation (MAD), mean posterior
predictive variance (AVE VAR), median posterior predictive standard
deviation (MED SD) and coverage probability of 95\% intervals
(COV).
All values are multiplied by 100}\label{tmse}
\begin{tabular*}{\tablewidth}{@{\extracolsep{\fill}}lccccc@{}}
\hline
& \multicolumn{5}{c@{}}{$\bolds{M}$}\\[-4pt]
& \multicolumn{5}{c@{}}{\hrulefill}\\
& \textbf{1} & \textbf{2} & \textbf{3} & \textbf{4} & \textbf{5} \\
\hline
MSE & 18.9 & 18.6 & 18.3 & 18.2 & 17.9 \\
MAD & 23.5 & 22.7 & 22.2 & 22.0 & 21.4 \\
AVE VAR & 18.3 & 17.0 & 16.7 & 16.4 & 16.7 \\
MED SD & 41.9 & 40.0 & 39.2 & 38.8 & 38.2 \\
COV & 95.7 & 95.5 & 95.3 & 95.3 & 95.2 \\
\hline
\end{tabular*}
\end{table}

\begin{figure}

\includegraphics{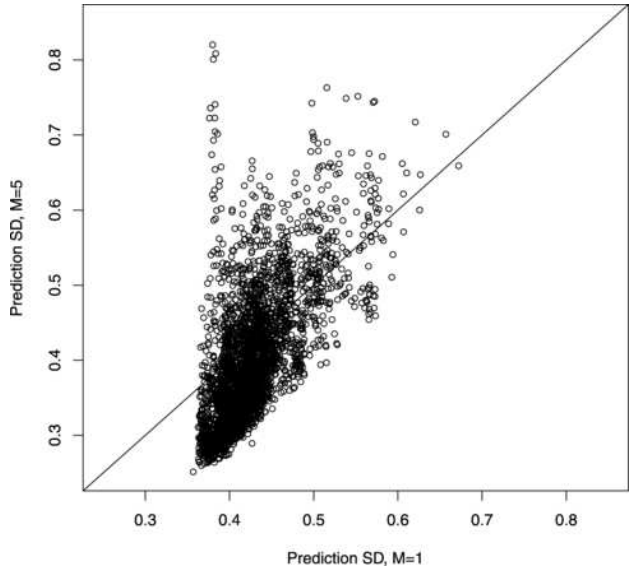}

\caption{Posterior predictive standard deviation for the
observations in the validation set for the stationary model with $M=1$
compared to the nonstationary model with $M=5$.}\label{fsd}
\vspace*{-3pt}
\end{figure}

The largest effect of nonstationary is in the measures of prediction
uncertainty. Figure \ref{fsd} plots the prediction standard deviation
for the observations in the validation set for the stationary model
with $M=1$ and the nonstationary model with $M=5$. The
standard
deviation is smaller for the nonstationary model for 72\% of the
observations, and varies far more across observations for the
nonstationary model (roughly from 0.25 to 0.80) compared to the
stationary model (roughly 0.35 to 0.65). To show that the conditional
coverage remains valid for both models, we separated the validation set
into five equally sized groups based on the ratio of standard
deviations from the models with $M=5$ to $M=1$. The coverage of 95\%
intervals in these five groups (from smallest to largest relative
variance) is 0.97, 0.97, 0.96, 0.96 and 0.93 for the stationary model
and 0.93, 0.96, 0.94, 0.97 and 0.96 for the nonstationary model.

%

%
\begin{table}
\tabcolsep=0pt
\caption{Summary of the model with $M=5$ components. The
remaining columns give the posterior means (95\% intervals) for the
mean effects $\beta_k$, the relative variance $(\Delta_k(0,0))$, the
relative spatial correlation at lag 100 km $({\tilde\Delta}_k(100,0))$,
and the relative temporal correlation at lag 2 days $({\tilde\Delta
}_k(0,2))$. $\beta_k$, $\Delta_k(0,0)$, ${\tilde\Delta}_k(100,0)$, and
${\tilde\Delta}_k(0,2)$ are scaled to represent the effect of a two
standard deviation increase in the predictor}\label{teffects}
{\fontsize{8.5pt}{11pt}\selectfont{
\begin{tabular*}{\tablewidth}{@{\extracolsep{4in minus 4in}}llccc@{}}
\hline
& \multicolumn{1}{c}{\textbf{Mean}}
& \multicolumn{1}{c}{\textbf{Variance}}
& \multicolumn{1}{c}{\textbf{Spatial cor.}}
& \multicolumn{1}{c@{}}{\textbf{Temporal cor.}}\\
& \multicolumn{1}{c}{$\bolds{\beta_k}$}
& \multicolumn{1}{c}{$\bolds{\Delta_k(0,0)}$}
& \multicolumn{1}{c}{$\bolds{{\tilde\Delta}_k(100,0)}$}
& \multicolumn{1}{c@{}}{$\bolds{{\tilde\Delta}_k(0,2)}$}\\
\hline
Temperature (F) & \hphantom{$-$}0.333 ($0.331, 0.358$) & 1.09 ($1.06, 1.12$) & 0.88
($0.86, 0.90$) & 1.09 ($1.03, 1.16$) \\
Wind speed (m/s) & $-$0.028 ($-0.037, -0.019$) & 0.96 ($0.94, 0.97$) & 1.05
($1.04, 1.06$) & 0.97 ($0.94, 1.00$) \\
Cloud cover (\%) & $-$0.154 ($-0.173, -0.134$) & 1.12 ($1.07, 1.17$) & 1.05
($1.03, 1.06$) & 0.57 ($0.51, 0.64$) \\
Elevation (ft) & \hphantom{$-$}0.115 ($0.078, 0.183$) & 0.98 ($0.96, 1.01$) & 1.10
($1.09, 1.11$) & 1.30 ($1.24, 1.37$) \\
Urban & \hphantom{$-$}0.007 ($-0.020, 0.035$) & 1.00 ($0.98, 1.02$) & 0.95 ($0.93,
0.96$) & 0.99 ($0.96, 1.02$) \\
Rural & \hphantom{$-$}0.045 ($0.010, 0.066$) & 0.57 ($0.43, 0.77$) & 0.94 ($0.90,
0.97$) & 1.17 ($1.02, 1.37$) \\
Year & \hphantom{$-$}0.004 ($-0.001, 0.008$) & 1.01 ($1.00, 1.02$) & 1.03 ($1.02,
1.03$) & 0.95 ($0.93, 0.97$) \\
Longitude & \hphantom{$-$}0.096 ($0.018, 0.187$) & 0.99 ($0.94, 1.03$) & 1.12
($1.10, 1.13$) & 1.53 ($1.43, 1.62$) \\
Latitude & \hphantom{$-$}0.185 ($0.089, 0.251$) & 0.70 ($0.67, 0.74$) & 1.05 ($1.03,
1.07$) & 0.48 ($0.40, 0.55$) \\
Temp$^2$ & \hphantom{$-$}0.023 ($0.014, 0.032$) & -- & -- & -- \\
WS$^2$ & \hphantom{$-$}0.004 ($0.002, 0.005$) & -- & -- & -- \\
CC$^2$ & $-$0.020 ($-0.029, -0.010$) & -- & -- & -- \\
Temp${}\times{}$WS & $-$0.005 ($-0.013, 0.003$) & -- & -- & -- \\
Temp${}\times{}$CC & \hphantom{$-$}0.055 ($0.044, 0.068$) & -- & -- & -- \\
WS${}\times{}$CC & \hphantom{$-$}0.004 ($-0.003, 0.012$) & -- & -- & --\\
\hline
\end{tabular*}}}
\end{table}

\begin{figure}

\includegraphics{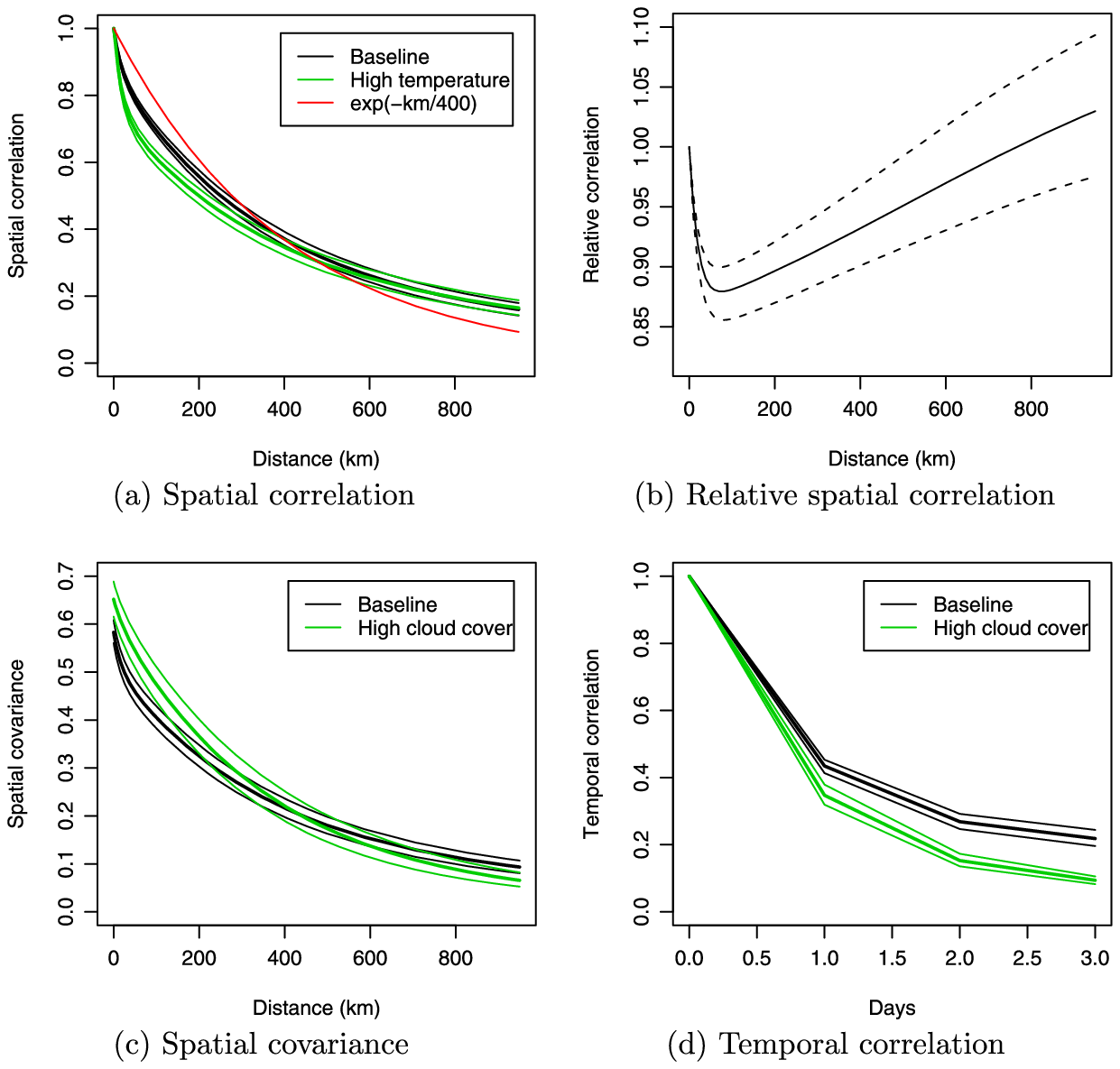}

\caption{Posterior mean (thick lines) and 95\% intervals
(thin lines) of the spatiotemporal correlation (\protect\ref{cov1})
for various
combinations of the covariates. ``Baseline'' assumes that all
covariates are zero (the mean after standardization) for both
observations. The other plots assume that all covariates are zero with
the exception of one covariate, which equals two standard deviation
units above the mean. Panel \textup{(b)} plots the posterior of ratio of the
spatial correlations under high temperature and baseline conditions
plotted in panel \textup{(a)}. The spatial correlation is plotted as a function
of spatial distance $h_s$ with temporal distance $h_t=0$, and vice
versa.}\label{fcovcor}
\end{figure}

Table \ref{teffects} and Figure \ref{fcovcor} summarize the covariate
effects on the mean and spatiotemporal correlation for the full data
set with $M=5$. The mean trend accounts for most of the variability in
square root ozone: though the sample variance of the observations is
1.61 ppb, the posterior means of the spatial effects $\delta(\bs)$ have
variance 0.09 ppb. The statistical significance of the linear and
quadratic temperature terms and the positive effect of temperature on
variance are consistent with findings of \citet{nhm2010} and
others that
ozone concentration is a monotone increasing nonlinear function of
temperature, and ozone variance increases with the mean. It is
reasonable that spatial correlation decreases as temperature increases
due to the fact that when the solar radiation is conducive to the
chemical reactions that produce ozone, that production is a function of
local emissions, and highly nonlinear in NOx emissions, which vary over
space. Similarly, it is reasonable that spatial correlation at short
spatial lags increases with wind speed because wind facilitates
transport of ozone and its precursors.

%

As discussed in Section \ref{svariogram}, the relationship between
cloud cover and ozone is quite complex. We find that cloud cover is
negatively associated with the mean and temporal correlation, and
positively associated with variance and\vadjust{\goodbreak} spatial correlation. As
expected, mean ozone decreases and spatial correlation increases with
cloud cover since ozone levels drop near low, heterogenous background
levels in the absence of solar radiation. A~possible explanation for
low variance and high temporal autocorrelation for sunny days is the
common southeastern summertime meteorological regime called the
``Bermuda high,'' which is characterized by sunny skies and high
atmospheric pressure indicative of a lower atmospheric boundary layer.
The lowered ceiling combined with low wind speed effectively reduce the
volume in which emissions interact, which, combined with high solar
radiation, creates a simmering cauldron of ozone production. Because
the Bermuda high persists over several days and spans regions greater
than or equal to the size of our spatial domain, ozone production is
high everywhere, so that the variability is lower and the temporal
correlation is higher.

Figure \ref{fcovcor} plots the estimated spatial and temporal
covariance for several combinations of the covariates. Figures \ref
{fcovcor}(a) and \ref{fcovcor}(b) show that the estimated spatial
correlation is lower for spatial lags less than 100 km for hot days, and
that temperature is less relevant at larger distances. This plot also
shows the mixture of exponential correlation functions gives a
correlation that is significantly different than a simple exponential
correlation. The mixture correlation function drops more quickly near
the origin and has a~heavier\vadjust{\goodbreak} tail than an exponential correlation.
Cloud cover also affects both the spatial covariance and temporal
autocorrelation. Figure \ref{fcovcor}(c) shows that the variance is
higher on cloudy days, but the covariance has smaller spatial range.
Also, the temporal correlation in Figure \ref{fcovcor}(d) is higher for
lags one, two and three for sunny days.

\begin{figure}

\includegraphics{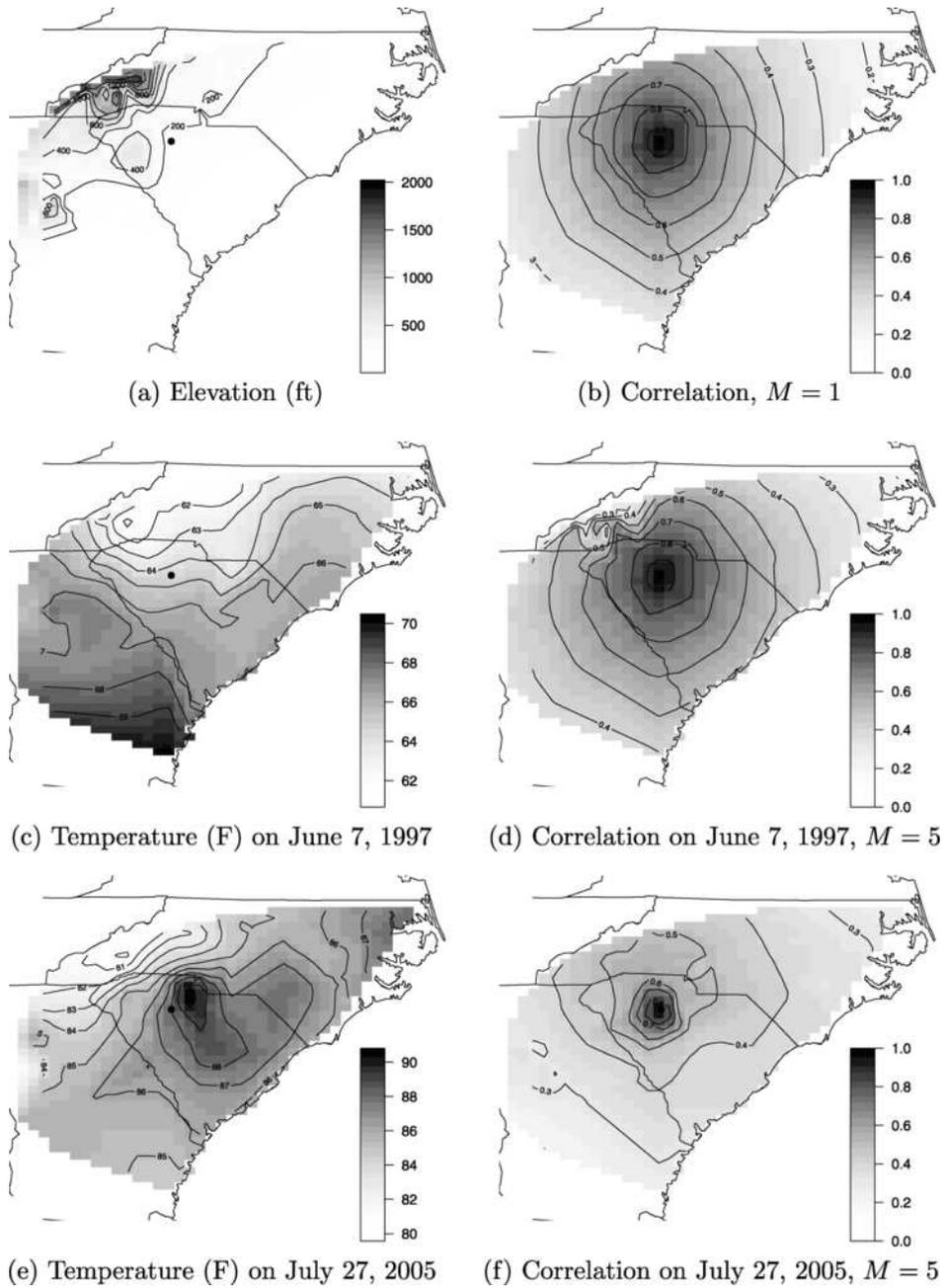}

\caption{Data and spatial correlation estimates for two days
for stationary ($M=1$) and nonstationary ($M=5$) models. Panels \textup{(b)},
\textup{(d)} and \textup{(f)} plot the posterior mean of the correlation between the
point marked with a dot and the remaining sites.}\label{fcorest}
\end{figure}

Figure \ref{fcorest} compares the posterior mean of the stationary
one-component model to that of the nonstationary five-component model,
%
%
and shows the relationship\vadjust{\goodbreak} between the spatial covariance of the latter
model with temperature and elevation. Figure \ref{fcorest}(b) shows the
exponential decay in correlation with increasing distance from the
marked site for the stationary model; this correlation function is the
same for the two days under consideration, June~7, 1997, and July 27,
2005, which have the minimum and maximum temperatures at the marked
site. The temperature contours for those days are plotted in Figures
\ref{fcorest}(c) and~\ref{fcorest}(e), and elevation contours are plotted
in Figure~\ref{fcorest}(a). The spatial correlation contours in the
northwest of Figure~\ref{fcorest}(d) show the negative effect of
elevation on spatial correlation. The July~27, 2005 position of the
maximum temperature peak over the marked site clearly shows the effect
of temperature on the steepness of the decline in correlation at short
versus long lags seen earlier in Figure \ref{fcovcor}(a). The effect of
elevation on correlation is dwarfed by the effect of temperature,
likely due to the positioning of the temperature peak over the marked
site combined with the magnitude of the temperature at that peak.

\section{Discussion}\label{sdisc}

In this paper we present a class of spatiotemporal covariance functions
that allows the covariance to depend on environmental conditions
described by known covariates. Although fitting this, and other
sophisticated spatiotemporal models, likely requires expertise in
spatial statistics and computing methods, the method produces
interpretable summaries of the effect of each covariate on the mean,
variance, and spatial and temporal ranges. For the southeastern US
ozone data, we find our nonstationary analysis improves prediction
error, reduces prediction variance, and achieves the desired coverage
probabilities, while identifying several interesting covariate effects
on both the mean and covariance.

Our covariance model assumes that all nonstationarity can be explained
by the spatial covariates. However, in some cases a more flexible model
would be useful. One approach would be to add more pure functions of
space and time as covariates in the covariance to capture
nonstationarity. An even more flexible model would take the weights to
be Gaussian processes, possibly with means that depend on the
covariates, to allow the weights to vary smoothly through the spatial
domain while still making use of the covariate
information.

\section*{Acknowledgments}
The authors wish to thank the Editor, Associate Editor and referees for
their helpful comments which greatly improved the manuscript.


%

%
\printaddresses

\end{document}